\documentstyle[psfig,epsf,conf-X,10pt]{article}
\begin{document} 
\small
\heading{%
%
Simulated Cluster Archive: A Computational Catalog of X-Ray Clusters
in a $\Lambda$CDM Universe
}
\par\medskip\noindent
\author{%
Michael Norman$^{1,2}$, Greg Daues$^{1}$, Erik Nelson$^{2}$,
Chris Loken$^{3}$, Jack Burns$^{3}$,
Greg Bryan$^{4}$, Anatoly Klypin$^{5}$
}
\address{%
National Center for Supercomputing Applications, Urbana, IL 61801
}
\address{%
Astronomy Department, University of Illinois, Urbana, IL 61801
}
\address{%
Physics Department, University of Missouri, Columbia, MO 65211
}
\address{%
Physics Department, MIT, Cambridge, MA 01238
}
\address{%
Department of Astronomy, New Mexico State University, Las Cruces, NM 88003                       
}

\begin{abstract}
We have simulated the evolution of a large sample of X-ray clusters 
in a $\Lambda$CDM universe at high spatial resolution using adaptive 
mesh refinement and placed the results in an online archive for public
access. The Simulated Cluster Archive website {\tt sca.ncsa.uiuc.edu} 
provides tools for interactive 2D and 3D analysis of gas and dark
matter fields, X-ray and SZ imaging, and data export. We encourage 
community use and solicit their feedback.
\end{abstract}
\section{Why a Simulation Archive?}
The creation of public archives of high-value observational
data (e.g., NASA's HSEARC) has been a great boon to astronomical
research in the past decade. It has given rise to a new kind of
astronomer---the archival astronomer---who is free to check the
methods and results of the data's authors, as well as to pursue
independent, and sometimes novel, lines of inquiry. Archival
astronomy is expected to grow in scope and importance in the
coming decade. By creating this archive of simulated X-ray clusters, we
hope to extend this concept into the computational realm where simulations
are growing in size and complexity.

%
\section{Science Goals}  
Our goal is to produce large, statistical catalogs of $\sim$ 
100 clusters simulated at high resolution for two cases:
(1) with, and (2) without non-adiabatic physical processes,
in order to:
\vskip 2pt 
\noindent 
\null~~~$\bullet$ {understand the role of non-adiabatic
processes in X-ray clusters,}\\
\null~~~$\bullet$ {make definitive predictions of the
XLF evolution in both cases,}\\
\null~~~$\bullet$ {compare degree of cluster substructure
with observational samples,}\\
\null~~~$\bullet$ {determine frequency of cooling flows
as a function of $z$.}\\
%
%

We simulate a $\Lambda$CDM model with parameters $h=0.7,
\Omega_m=0.3, \Omega_b=0.026, \Omega_{\Lambda}=0.7, \sigma_8=0.928$.
The survey volume is $256 h^{-1}$ Mpc on a side.
We employ a new hydro+N-body code \cite{BN99} which uses
Adaptive Mesh Refinement to place high resolution grids where needed.
First, a survey calculation was performed with $256^3$ cells,
$128^3$ particles, and two levels of refinement everywhere to locate
the clusters in our sample. Then, each cluster is recomputed with
up to 7 levels of refinement within the cluster environment.
The DM mass resolution is $\sim 10^{10} h^{-1} M_{\odot}$; the L7
spatial resolution is $15.6 h^{-1}$ kpc.   

The adiabatic (control) sample is near completion
and being analyzed. Preliminary results focusing on the ten brightest
clusters have been reported in \cite{Burns99} and a second, more
extensive paper is in preparation.

\section{Archive Design and Tools}
AMR simulation data structures are hierarchical and complex, and 
require specialized software for their manipulation and
analysis \cite{Norman99}. One of the design goals of the SCA 
was to shield the user from the complexity (and size) of AMR simulation 
data. We accomplished this by developing the SCA as a workbench-style
system that lets users interact with the archived data over the Web. 
The user begins by selecting a cluster from a catalog list or
a 3D VRML map. The server then retrieves the raw AMR data from
NCSA's mass storage system. Once the data is on the SCA web server,
the user extracts from the AMR files particle data and field data
sampled to a uniform grid of user-specified size and resolution.
At this point, the user may export the extracted data as HDF files
for local analysis or use the suite of analysis tools provided
as a part of the SCA. The tools, which include 2D and 3D visualization
tools, X-ray and SZ imager, and graphing tool,
are implemented as thin Java applets in a client-
server model. More detail can be found in \cite{Daues99}.
 
\acknowledgements{This work was supported by NASA grant NAG5-7404}
\vfill
\end{document}